\documentclass[12pt]{iopart}

\usepackage{cancel}
\usepackage{graphicx}
\usepackage{dcolumn}
\usepackage{bm}
\usepackage{xcolor}

\newcommand{\be}{\begin{equation}}
\newcommand{\ee}{\end{equation}}
\newcommand{\la}{\langle}
\newcommand{\ra}{\rangle}

\newcommand{\lf}{\left}
\newcommand{\rg}{\right}
\newcommand{\eqref}[1]{(\ref{#1})}

\begin{document}

\title[Finite-energy L\'evy-type motion and heterogeneity]{Finite--energy L\'evy--type motion through heterogeneous ensemble of Brownian particles}

\author{Oleksii Yu. Sliusarenko}
\address{BCAM--Basque Center for Applied Mathematics, Alameda de Mazarredo 14, E--48009 Bilbao, Basque Country, Spain}
\author{Silvia Vitali}
\address{Department of Physics and Astronomy, Bologna University, Viale Berti Pichat 6/2,
I--40126 Bologna, Italy}
\author{Vittoria Sposini}
\address{Institute for Physics and Astronomy, University of Potsdam, Karl--Liebknecht--Strasse 24/25, D--14476 Potsdam--Golm, Germany}
\address{BCAM--Basque Center for Applied Mathematics, Alameda de Mazarredo 14, E--48009 Bilbao, Basque Country, Spain}
\author{Paolo Paradisi}
\address{ISTI--CNR Institute of Information Science and Technologies ``A. Faedo'', Via G. Moruzzi 1, I--56124 Pisa, ITALY}
\address{BCAM--Basque Center for Applied Mathematics, Alameda de Mazarredo 14, E--48009 Bilbao, Basque Country, Spain}
\author{Aleksei Chechkin}
\address{Institute for Physics and Astronomy, University of Potsdam, Karl--Liebknecht--Strasse 24/25, D--14476 Potsdam-Golm, Germany}
\author{Gastone Castellani}
\address{Department of Physics and Astronomy, Bologna University, Viale Berti Pichat 6/2,
I--40126 Bologna, Italy}
\author{Gianni Pagnini}
\address{BCAM--Basque Center for Applied Mathematics, Alameda de Mazarredo 14, E--48009 Bilbao, Basque Country, Spain}
\address{Ikerbasque--Basque Foundation for Science, Calle de Mar\'ia D\'iaz de Haro 3, E--48013 Bilbao, Basque Country, Spain}


\vspace{.5cm}
\hspace{1.6cm} E-mail: paolo.paradisi@cnr.it (Paolo Paradisi)

\hspace{1.6cm} E-mail: gpagnini@bcamath.org (Gianni Pagnini)

\date{\today}

\newpage

\begin{abstract}
Complex systems are known to display anomalous diffusion, whose signature is a
space/time scaling $x\sim t^\delta$ with $\delta \ne 1/2$ in the Probability Density Function (PDF).
Anomalous diffusion can emerge jointly with both Gaussian, e.g., fractional Brownian
motion, and power-law decaying distributions, e.g., L\'evy Flights (LFs) or L\'evy Walks
(LWs).
LFs get anomalous scaling, but also infinite position variance and, being jumps of any size
allowed even at short times, also infinite energy and discontinuous velocity.
LWs are based on random trapping events, resemble a L\'evy-type
power-law distribution that is truncated in the large displacement range and have finite
moments, finite energy and discontinuous velocity.
However, both LFs and LWs cannot describe friction-diffusion processes and do not
take into account the role of strong heterogeneity in many complex systems, such as
biological transport in the crowded cell environment.
%
%
We propose and discuss a model describing a Heterogeneous
Ensemble of Brownian Particles (HEBP) based on a linear Langevin equation.
We show that, for proper distributions of relaxation time and velocity
diffusivity, the HEBP displays features similar to LWs, in particular
power-law decaying PDF, long-range correlations and anomalous diffusion,
at the same time keeping finite position moments and finite energy.
The main differences between the HEBP model and two LWs are investigated, finding that,
even if the PDFs are similar, they differ in three main aspects:
(i)
LWs are biscaling, while HEBP is monoscaling; 
(ii)
a transition from anomalous ($\delta \ne 1/2$) to normal ($\delta = 1/2$) diffusion in
the long-time regime;
(iii)
the power-law index of the position PDF and the space/time
diffusion scaling are independent in the HEBP, while they both depend on the scaling of
the inter-event time PDF in LWs.
The HEBP model is derived from a friction-diffusion process, it has
finite energy and it satisfies the fluctuation-dissipation theorem.
%
%
%
%
%
%
%
%
%
\end{abstract}


\vspace{1.cm}

\maketitle

%

\newpage

\section{\label{sec:intro}Introduction}

\noindent
Diffusion and transport
play a central role in internal dynamical processes of many complex systems
and often represent their main drivings.
As an example, efficiency in the transport of chemicals and particles
affects reaction rates through the probability that two or more reacting
molecules
``meet'' each other \cite{rice1985,dorsaz_prl10}.
%
%
%
%
%
Another example is given by turbulent diffusion that is one of the main
mechanisms, the other one is advection by large-scale motions, driving the
dispersion of contaminants and pollutants (gas, aerosol particles, dust, seeds)
in the atmosphere \cite{kaplan1993,paradisi_pa01,paradisi_pceb01,paradisi_npg12,cheng_csb2014,goulart_pa17}.

\noindent
The first observation concerning diffusive motion of particles in fluids dates
back to a period between the 18th and 19th centuries
\cite{ingenhousz1784,bywater1819,brown_pm1828,brown_pm29}
(see, e.g., \cite{abbott_ieee-te1996} for a interesting historical perspective).
%
%
The so-called {\it normal, Brownian} or {\it standard diffusion} was firstly
observed. This is defined by two conditions: (i) Mean-Squared Displacement
(MSD) grows linearly in time: $\la x^2 \ra = 2 D t$
and (ii) Probability Density Function (PDF) of particle displacements
is Gaussian \footnote{
  Both conditions follow from the well-known Central Limit Theorem, which states the emergence of a Gaussian random variable from the sum of many random
  contributions that are both finite-size (i.e., finite variance) and
  statistically independent (i.e., uncorrelated).
}.

\noindent
Normal diffusion has been historically the first observed and theoretically
investigated diffusive motion as it emerges in non-complex, i.e., not
self-organized systems, thus without coherent structures or complex
heterogeneous conditions that can affect diffusive motion in a non-trivial way
(e.g., by introducing long-range correlations).
This was the condition usually observed in the kind of experiments made by
Brown, Perrin and others,  usually a still liquid or a gas at equilibrium
(\cite{ingenhousz1784,bywater1819,brown_pm1828,perrin_cr1908,perrin_acp1909}.
%

\noindent
On the contrary, when complex systems are considered, that is, characterized by the
emergence of self-organized states, i.e., coherent large-scale, long-time
lasting, structures,
%
%
deviations from the linear time-dependence of the variance are
typically observed \cite{metzler_etal-physrep-2000,paradisi_csf15_preface,paradisi_complex_csf15}:
\be
\la x^2 \ra = 2 D_\phi t^\phi = 2 D_H t^{2 H}\ \  {\rm with}\ \  \phi \ne 1\  
(H \ne 1/2 \ )\ .
\label{anom_diff}
\ee
$H$ is the Hurst exponent or {\it second moment scaling} and  $H=1/2$
identifies the normal diffusion scaling.
This condition is known as {\it anomalous diffusion}
\cite{metzler_etal-physrep-2000,metzler_etal-jpa-2004,klages_anomalous_2008}.
%
%
It is worth noting that Eq. (\ref{anom_diff}) shows that not only the global
efficiency of the
diffusion, but also the particular kind of transport is a crucial property,
being the first one measured by the generalized position diffusivity $D_\phi$ and the
second one encoded in the diffusion scaling $\phi = 2 H$.
%
%
%

\noindent
The first observed anomalous diffusion dates back to the Richardson's t-cubed
law for the relative particle diffusion in turbulence, which was already
reported in 1926 \cite{richardson_prsa1926}.
Another historically important example comes from the motion of charge carriers
in amorphous semiconductors, which was extensively studied by Montroll and
co-workers
(see, e.g., \cite{montroll1964,montroll_etal-jmp-1965,scher_etal-prb-1975}).
%
%
%
In the last three decades, 
the number of complex self-organized systems displaying
%
%
anomalous diffusion has increased very rapidly
\cite{solomon_prl93,kraichnan_prl94,claused_pa97,venkataramani_pd98,periasamy_bj98,delcastillo-negrete_pp04,furstenberg_csf15}.
In particular, in the field of biological transport many new experimental findings
are being published every year
\cite{tolicnorrelykke_etal-prl-2004,golding_etal-prl-2006,gal_experimental_2010,barkai_etal-phystod-2012,manzo_etal-rpp-2015,ariel2015swarming,he_nc16}
and this is attracting 
a great interest in the scientific community of theoreticians with many models
being proposed and compared with data
\cite{jeon_etal-prl-2011,hofling_etal-rpphys-2013,manzo_etal-prx-2015,molina_etal-pre-2016}.
In particular, many papers are being devoted to model the random diffusive motion
of macromolecules in the cell cytoplasm and membrane 
\cite{manzo_etal-prx-2015,reverey_etal-scirep-2015,molina_etal-pre-2016,metzler_etal-bba-2016,jeon_prx16},
or in artificial {\it in vitro} environments \cite{dix_arb08,nawrocki_jpcb17}, such as a
mixture of water, proteins and lipids, used to mimick and investigate mechanisms
occurring in biology (e.g., trapping of proteins by simultaneously forming lipid
vesicles) \cite{luisi_bbab09,luisi_cbc10,paradisi_bmcsb15}.

%
%
%

\noindent
The first proposed model for anomalous diffusion is the Continuous Time Random
Walk (CTRW), which was introduced and extensively studied and applied by Montroll
and co-workers \cite{montroll1964,montroll_etal-jmp-1965,scher_etal-prb-1975}
(see \cite{weiss1983} for a review).
Its very first version is a random walk with statistically independent random
jumps and random times that are also decoupled with each other.
Random times, also called Waiting Times (WTs), describe a trapping mechanism
due to a sequence of potential wells
\cite{bouchaud_etal-physrep-1990,bouchaud-jpf-1992}, thus this
particular CTRW model can describe only subdiffusion ($\phi<1$).
The WT is the intermediate long time between two crucial short-time events,
each one given by the escape from a given well and the jump into another one,
thus CTRW is essentially driven by the sequence of WTs, described by a
renewal point process
\cite{bianco_cpl07,paradisi_epjst09,paradisi_npg12,paradisi_romp12,paradisi_csf15_pandora,paradisi_springer2017}.
Several CTRW models have been introduced and investigated, but the
subdiffusive CTRW remains probably the most studied and applied one, with the
exception of
so-called L\'evy Walk (LW) model, which is a CTRW whose jumps and WTs are
coupled \cite{shlesinger_random_1982,shlesinger_pa86,zaburdaev_rmp15}.
Unlike the subdiffusive uncoupled CTRW, LWs
can indeed reproduce superdiffusive behavior.
CTRW represents an important modeling approach extensively applied to many
complex systems, such as biological transport (see, e.g.,
\cite{burov_etal-pccp-2011,reverey_etal-scirep-2015,metzler_etal-bba-2016} for subdiffusive CTRWs and \cite{de_jager_levy_2011,ariel2015swarming} for  L\'evy Walks and search
strategies).
%
%
%
Other models do not consider the existence of crucial jump events, while
explicitly including the long-range correlations of the process in the
dynamical equations. This is the case of Fractional Brownian Motion (FBM)
\cite{mandelbrot_etal-siamrev-1968,biagini2008} 
and of a viscoelastic model such as the Generalized Langevin Equation (GLE)
\cite{heppe-jfm-1998,goychuk_acp13,goychuk_po14,stella_prb14}, which are both
essentially based on Gaussian stochastic processes.
%
%

\vspace{.3cm}
\noindent
{\it Anomalous diffusion from heterogeneity}

\vspace{.05cm}
\noindent
CTRW and FBM had some success in applications to biological transport.
However, each of these models does not seem able to take into account all the
observed statistical features of transport
\cite{massignan_etal-prl-2014,manzo_etal-prx-2015,molina_etal-pre-2016},
so that a unified reasonable physical picture describing experimental data
does not yet exist.
A new direction in theoretical modeling recently emerging in the
scientific community comes from a quite simple observation:
diffusion in biological environments like the cell cytoplasm or
membrane is mainly affected by the very complex heterogeneity, the crowding
and the presence of different kinds of structures (e.g., cytoskeleton).
For this reason, a great attention on the role of heterogeneous environments in
anomalous diffusion is rapidly increasing and
an intense debate is raising in the scientific community,
especially in the context of anomalous biological transport \cite{hofling_etal-rpphys-2013,manzo_etal-prx-2015,molina_etal-pre-2016,lanoiselee_jpamt18,sposini_njp18}.
%
%

\noindent
Due to the above reasons, in very recent years the proposal of
Heterogeneous Diffusivity Models (HDMs) is taking momentum in the
scientific community
\cite{cherstvy_etal-njp-2013,jeon_etal-pccp-2014,cherstvy_etal-pccp-2016}.
Superstatistics is probably the first model of anomalous diffusion that is
based on the idea of a heterogeneous environment
\cite{beck_prl01,beck_pa03,vanderstraeten_pre09},
%
%
%
but a great attention is nowadays focused towards other approaches
trying to go beyond superstatistics. In particular, Diffusing
Diffusivity Models (DDMs) are being proposed
and studied in very recent literature
\cite{chubynsky_etal-prl-2014,chechkin_prx17,jain_jcs17,lanoiselee_jpamt18,sposini_njp18}.
In DDMs an additional stochastic equation is introduced to describe the
position diffusivity.
A similar but different approach, included into the class of HDMs,
follows from the very first idea of Schneider's grey Brownian Motion
(gBM)
\cite{schneider_1990,schneider_1992}.
In the gBM a random amplitude multiplying a Gaussian process, usually the FBM, is
introduced.  
This amplitude characterizes the motion of  single trajectories, so that the diffusion
properties of the ensemble are affected by the amplitude distribution.
In particular, gBM is associated with a Mainardi distribution of the amplitude
\cite{mainardi_etal-ijde-2010,pagnini-fcaa-2013} and, in this
case, the displacement PDF satisfies a time-fractional diffusion equation
\cite{gorenflo_etal-nd-2002,paradisi_caim15, sandev_fcaa18}.

%
%
In the last decade the gBM model was extended to the generalized grey Brownian
Motion (ggBM)
\cite{mura-phd-2008,mura_etal-pa-2008,mura_etal-jpa-2008,mura_etal-itsf-2009,pagnini_etal-ijsa-2012,pagnini_etal-ptrsa-2013}.
The ggBM was shown to satisfy the Erd\'elyi--Kober fractional diffusion equation
\cite{pagnini-fcaa-2012}, which includes the time-fractional diffusion equation,
describing the gBM
distribution, as a particular case.
A further generalization is given by the ggBM-like model discussed by Pagnini
and Paradisi, 2016 \cite{pagnini_etal-fcaa-2016}, which was proven
to satisfy the space-time fractional diffusion equation
\cite{gorenflo_etal-cp-2002,gorenflo_etal-pa-2002,mainardi_etal-fcaa-2001}
regardless of the particular Gaussian process describing single
trajectory's dynamics \footnote{
  It is worth noting that, similarly to ggBM, the model discussed in Ref.
  \cite{pagnini_etal-fcaa-2016} reduces to
  time-fractional diffusion for a proper choice of parameters.
}.
For this reason, this class of ggBM-like models is here denoted as Randomly Scaled
Gaussian Processes (RSGPs), as it extends the ggBM not only to much more general
space-time fractional diffusion, but it also includes whatever Gaussian process as the
process driving single trajectory dynamics. 
The DDM approach has been recently compared
against a ggBm-like approach with a random scale governed by the same
stochastic differential equation \cite{sposini_njp18}.
The potential application of ggBM-like models to biological transport
was discussed by showing that the behavior of a set of different statistical
indices are qualitatively accounted for by this kind of modeling approach
\cite{molina_etal-pre-2016}.
However, to our knowledge, DDMs, gBM and ggBM-like models do not directly describe the
particle velocity's dynamics and, thus, the role of friction and velocity
diffusivity are not explicitly taken into account.
%
%

\vspace{.3cm}
\noindent
{\it Heterogeneous ensemble of Brownian particles and RSGPs}

\vspace{.05cm}
\noindent
To overcome this limitation, the dynamics of a Heterogeneous Ensemble of Brownian
Particles (HEBP) have been recently investigated by Vitali et al., 2018
\cite{vitali_jrsi18,dovidio_spta18}, where 
a stochastic model that takes explicitly into account
the heterogeneity is derived,
but this model does not belong neither to the class of DDMs nor to that of HDMs.
It is instead based on a linear Langevin equation for a friction-diffusion
(i.e., Ornstein--Uhlenbeck) process that describe the velocity dynamics.
A population of relaxation time and velocity diffusivity parameters are then considered,
that is, mathematically treated as random variables whose statistical
distributions are derived by imposing the emergence of anomalous diffusion,
long-range correlations and power-law decay in the position distribution of the particle
ensemble (see following section for model details).
This means to assume that particles in the ensemble follow different dynamics
depending on the different physical parameters.
Due to linearity, this model is easily recognized to be equivalent to a RSGP
for both position and velocity:
\be
x(t) = \sqrt{2D}\, x_G(t)\ ;\quad v(t) = \sqrt{2D}\, v_G(t)\ ,
\label{rsgp}
\ee
being $x_G(t)$ and $v_G(t)$ proper Gaussian processes and $D$ a random velocity
diffusivity (see \ref{rsgp-ggou}).
In the RSGP model \eqref{rsgp} the single trajectory is still described by a Gaussian
process,
but this is no more a FBM. It instead follows from the joint effect of the different
relaxation time scales $\tau$.
For proper distribution of $\tau$, this causes the emergence of long-range
correlations and anomalous, but still Gaussian,
diffusion with scaling $\phi = 2 H \ne 1$.
Conversely, the non-Gaussianity in the Probability Density Function (PDF) of the position
is related to inhomogeneities in the velocity diffusivity $D$ \cite{vitali_jrsi18}.
%
%
An interesting point deserving attention is that the HEBP/RSGP model proposed by
Vitali et al., 2018 \cite{vitali_jrsi18} has a clear physical meaning as it
describes the dynamics of an ensemble of Brownian particles
with heterogeneous physical properties and moving in a viscous medium
in thermal equilibrium, thus giving a well-posed physical basis to ggBM and
ggBM-like processes (i.e., RSGPs) \cite{dovidio_spta18}.
%
%

\vspace{.3cm}
\noindent
{\it The problem of infinite energy}

\vspace{.05cm}
\noindent
As known, anomalous diffusion is often observed jointly with non-Gaussian PDFs
displaying slow power-law decaying tail: $p(x,t) \sim 1/x^{1+\alpha}$ with
$0 < \alpha < 2$.
For this kind of non-Gaussian PDFs, the HEBP model developed by Vitali et al., 2018
\cite{vitali_jrsi18} share with other anomalous diffusion processes, such as
L\'evy flights, the problem of an infinite variance, thus formally allowing for
a physically meaningless infinite energy in the system.
Furthermore, this does not allow to have
a fluctuation-dissipation theorem for the equilbrium velocity PDF in a
stationary thermal bath.

\noindent
To overcome this limitation, while remaining in the framework of
heterogeneity-driven anomalous diffusion, we here discuss a simple and natural
modification of
the HEBP model proposed in Vitali et al., 2018 \cite{vitali_jrsi18} and show that this
modification is sufficient to get a physically meaningful model, at the same
time being able to reproduce behaviors similar to those of other anomalous
diffusion processes, in particular L\'evy Walk (LW) models.
In fact, similarly to LW, our proposed model is shown to display a 
power-law decay of the distribution for intermediate values of the position,
at the same time keeping the finiteness of moments and, thus, of energy due
to an exponential cut-off in the distribution tails.
However, in spite of its finite energy, LW cannot describe a friction-diffusion process and,
thus, fluctuation-dissipation theorem does not apply.

%
%

\vspace{.2cm}
\noindent
The paper is organized as follows.
In Section \ref{sec:descr} we discuss the stochastic model for the HEBP and we show
numerical simulations of the model.
In particular,  an anomalous-to-normal transition is shown to occur in Section
\ref{anom-to-normal}.
%
%
In Section \ref{sec:lw} the comparison of our model with
two different LWs is carried out. Finally, in Section \ref{discuss}
some discussions and final remarks are sketched.

\section{\label{sec:descr} Heterogeneous ensemble of Brownian particles}

\subsection{Preliminary considerations}
\label{model_1}

\noindent
Starting from the Langevin equations associated to each Brownian particle of
the ensemble,
the HEBP approach leads to anomalous diffusion with uncorrelated white noise.
Thus, HEBP models are substantially different from approaches based on the generalized
Langevin equation or on Langevin equations with colored noises and,
in general, on noises with long-range spatiotemporal correlations with even
"anomalous" thermodynamics \cite{gheorghiu_etal-pnas-2004}.
In HEBP models anomalous diffusion emerges as a consequence of heterogeneity in the
particle ensemble, while classical thermodynamics still hold.
Heterogeneity is then responsible for long-range correlations, in agreement
with approaches based on polydispersity \cite{gheorghiu_etal-pnas-2004}.
In particular, in the present approach anomalous behavior is displayed during
an intermediate asymptotic transient regime in the Barenblatt's sense
\cite{barenblatt-1979}, thus requiring an underdamped (white noise) Langevin
approach.
These last two features of anomalous diffusion are consistent 
with the findings in the case of the underdamped scaled Brownian motion
\cite{bodrova_etal-sr-2016}, 
and, implicitly, with the role of friction when a complex potential is applied
\cite{sancho_etal-prl-2004}.

\noindent
HEBP models are compared in literature with similar approaches based on fluctuating
friction 
\cite{rozenfeld_etal-pla-1998,luczka_etal-pa-2000,luczka_etal-appb-2004},
fluctuating mass \cite{ausloos_etal-pre-2006} 
and with the already cited DDM approach \cite{chechkin_prx17,sposini_njp18}.
%
%
Further approaches using a population of the involved parameters were proposed on the basis of a Gaussian processes, see for example the 
Markovian continuous time random walk model with a population of time-scales \cite{pagnini-pa-2014} or
the ggBM \cite{mura_etal-jpa-2008,molina_etal-pre-2016} that actually is the fBm with a population of length-scales.
Interestingly, approaches based on fluctuating friction or mass, as such as HEBP models,
are underdamped processes on the contrary of the DDMs and ggBm-like processes
\cite{mura_etal-jpa-2008,pagnini_etal-fcaa-2016,sposini_njp18},
that are overdamped.
In systems displaying anomalous diffusion, underdamped processes were shown to be a
preferable approach \cite{bodrova_etal-sr-2016}.
All these approaches take into account a distributed parameter and, 
then, they can be linked to superstatistics \cite{beck_pa03}.
The discussion of the present approach within
the idea of superstatistics is reported in Section \ref{superstat}.

\vspace{.2cm}
\noindent
In the HEBP model introduced here the ensemble of particles differ in their
density (mass divided by volume).
%
The main difference with mentioned approaches is that in our formulation
fluctuations refer to differences among particles and not to changes in time.
%
Particles differ
in their mass $m$, in their friction coefficient $\gamma$
and in their noise amplitude $b$, related to velocity diffusivity through: $D=b/m^2$.
The fluctuation-dissipation theorem states $b=\kappa_{\rm{B}} T \, \gamma$, where
$\kappa_{\rm{B}}$ and $T$ are the Boltzmann constant and the temperature, respectively.
Then, the set of distributed independent parameters $\{m,\gamma,b\}$ 
reduces to the set $\{m,\gamma\}$.
Moreover,  by assuming that the present one-dimensional model is indeed a Cartesian
direction of a three-dimensional isotropic and spatially independent process, the friction
coefficient is given by the Stokes law  $\gamma=6 \pi \nu r$ where $\nu$ is the viscosity
of the medium and $r$ the radius of the Brownian particle. 
This means that 
by the combination of the fluctuation-dissipation theorem and the Stokes law  
the set of distributed independent parameters is $\{m,r\}$.
Considering the definitions $\tau=m/\gamma$ and $D=b/m^2$,
the particle density (mass divided by volume)
is approximately $3 m/(4\pi r^3) = 162 \, \pi^4 \nu^3 \, D^2 \tau^5$
and the differences among particles in terms of $\{m,r\}$
translate into differences in terms of $\{\tau,D\}$,
namely the ensemble of particles is characterized by a population of 
diffusivities $D$ and a population of relaxation times $\tau$.
%
%
In this framework,  we highlight that both the populations of masses and radii contribute
to the emergence of the anomalous scaling,
by means of the relaxation times $\tau=m/\gamma=m/(6\pi \nu r)$,
and to the shape of the resulting probability density functions of particle dispersion, 
by means of the diffusivity $D=\kappa_{\rm{B}} T \, 6 \pi \nu \, r/m^2$.

%
\subsection{Model description}
\label{model_2}

\noindent
We consider an ensemble of particles with heterogeneous physical parameters moving in
a viscous medium.
Each particle moves according to a linear Langevin equation for a friction-diffusion,
i.e., Ornstein--Uhlenbeck process:
\begin{eqnarray}
&&\frac{dx}{dt} = v\ ,
  \label{eq:kinematic} \\
  \ \nonumber \\
&&  \frac{dv}{dt} = -\frac{1}{\tau} v(t) + \sqrt{2 D}\, \xi(t) \ .
 \label{eq:langevin_new}
\end{eqnarray}
As anticipated in the previous Section \ref{model_1}, $\tau$ and $D$ are the
viscous relaxation time and the velocity diffusivity, respectively.
The fluctuation-dissipation theorem for the single particle in the HEBP is given by:
\be
\tau D\  = \frac{\kappa_{\rm B} T}{m}= \la v^2 | \tau, D \ra_{\rm eq}\ .
\label{fluct-diss_single}
\ee
In our HEBP model each single particle has a different pair of
parameters $(\tau, D)$, which meet fluctuation-dissipation relation \ref{fluct-diss_single}) and remain constant throughout the motion.
The complexity in the dynamics of the ensemble is mathematically
introduced by means of an effective randomness in the parameters of the Langevin
equation (\ref{eq:langevin_new}) and, thus, by means of proper statistical
distributions for $\tau$ and $D$.
Interestingly, for each pair $(\tau,D)$, every trajectory itself remains an ordinary
Brownian motion in a viscous medium, i.e.,
a Ornstein--Uhlenbeck process with Wiener (Gaussian) noise.
Thus, the overall complexity emerges as an average behavior of the entire
ensemble of particles, which individually move according to a standard
Ornstein--Uhlenbeck (Gaussian) process.
%
%

\noindent
In order to get both anomalous diffusion, due to long-range correlations, and
power-law behavior in the PDF $p(x,t)$, we choose the following distributions
of $\tau$ and $D$
\cite{vitali_jrsi18}:
\begin{eqnarray}
&&g(\tau) = \frac{\eta}{\Gamma(\eta)} \frac{1}{\tau}
L_{\eta}^{-\eta}\left(\frac{\tau}{\tau_*}\right)\ ,\quad 0<\eta<1\ ;
\label{eq:tau_distr}\\
&& f(D) = \frac{1}{D_*} L_{\alpha/2}^{-\alpha/2}\left(\frac{D}{D_*}\right)\ ,
\quad  1<\alpha<2\ ;
\label{eq:D_distr}
\end{eqnarray}
where $\Gamma(\cdot)$ is the Gamma function, $L_{\alpha}^{-\alpha}(\cdot)$ the
L\'evy extremal density with stability index $\alpha$
\cite{gnedenko-kolmogorov1954,feller1971}, $\tau_*$ is a reference time scale
and $D_*$ a reference scale for the velocity diffusivity \footnote{
As well known, the L\'evy's Generalized Central Limit Theorem states that
L\'evy stable densities $L_\alpha^\theta(x)$, with $\theta$ asymmetry parameter,
have a basin of attraction for a class of PDFs with slowly decaying power-law
tails: $p(x,t) \sim 1/|x|^{1+\alpha}$ with $0 < \alpha \le 2$.
As a consequence, the choice of $g(\tau)$ and $f(D)$ is a robust one and
is expected to apply in the context of complex systems, i.e., systems with
self-organizing features and emergent structures where power-law tails and
anomalous transport often emerge due to cooperative dynamics.
}.
In the following we set $\tau_*=D_*=1$.

\noindent
As well-known, with the exception of the Gaussian case ($\alpha=2$), the
Mean Square Displacement (MSD) of a L\'evy stable density $L_\alpha^\theta$
diverges and, for
$0 < \alpha \le 1$, also the mean $\la D \ra$ is infinite, which is
exactly the case of Eq. (\ref{eq:D_distr}) for the considered range of
parameters.
Conversely, the average relaxation time is finite and is given by:
$\la\tau\ra = \eta \tau_*/\Gamma(1/\eta)$.
$\eta$ is the model parameter determining the \textsl{space-time scaling} of
the diffusion process, while $\alpha$ affects the power-law decay emerging in
the position PDF $p(x,t)$.

\noindent
It was proved in Ref. \cite{vitali_jrsi18} that the process conditioned
to a particular value of $D$ is a Gaussian stochastic process with long-range
velocity correlation. In particular, the stationary correlation function and
the MSD are given by:
\begin{eqnarray}
&&R(t | D) = D\, \frac{\Gamma( 1 + \eta )}{\Gamma( 1 - \eta )}
\left( \frac{\Gamma(1/\eta)}{\eta} \right)^\eta
\la \tau \ra^{1+\eta} \, t^{-\eta}\ ,\quad 0<\eta<1\ ;
\label{corr_free} \\
\   \nonumber \\
&&\sigma_X^2(t | D) = \la x^2 | D \ra = 2 C D t^\phi\  , 
\quad  1 < \phi = 2-\eta < 2\ ;
\label{var_superdiff_1} \\
\  \nonumber \\
&&C = 
\frac{\Gamma(\eta+1)}{\Gamma(3-\eta )} 
\lf( \frac{\Gamma(1/\eta)}{\eta}\rg)^{\eta}\la\tau\ra^{1+\eta}\ ,
\label{var_superdiff_2}
\end{eqnarray}
thus resulting in a {\it superdiffusive scaling} regime.\\
The one-time marginal
PDF is a Gaussian density with zero mean and variance (MSD) $\sigma_X^2$:
${\cal G}(x,\sigma_X(t | D))$.
By averaging Eq. (\ref{fluct-diss_single}) over $\tau$, we get for any fixed, finite $D$
\cite{vitali_jrsi18}:
\be
\langle v^2 | D \rangle_{\rm eq} = \langle \tau \rangle D\ .
\label{cond_var}
\ee
%
It is worth noting that, similarly to the Fractional Brownian Motion (FBM),
this model belongs to the
class of Gaussian stochastic processes with stationary increments and
long-range correlations, as it can be seen from the power-law behavior in
Eqs. (\ref{corr_free}) and (\ref{var_superdiff_1}).
Thus, this is a valid alternative model, as it shares with FBM the emergence
of anomalous diffusion scaling, but with different velocity correlation function
derived within the well-defined physical framework of complex heterogeneity.

\noindent
When $D$ is distributed according to the PDF $f(D)$ given in Eq.
(\ref{eq:D_distr}), the probability of finding a particle
in $x$ at time $t$ is given by \cite{vitali_jrsi18}:
\begin{eqnarray}
p(x,t) &=& \int_0^\infty {\cal G}\left(x,\sigma_X(t|D)\right) \frac{1}{D_*}
L_{\alpha/2}^{-\alpha/2} \left(\frac{D}{D_*} \right) \mathrm{d}D =
\nonumber \\
\label{eq:space_fract}
&=& \frac{1}{\sqrt{ C D_* t^{\phi}} } L_\alpha^0
\left( \frac{x}{\sqrt{C D_* t^{\phi}} }\right)\ ,
\end{eqnarray}
with $\sigma_X(t | D)$ and $C$ given by Eqs. ~\eqref{var_superdiff_1} and
~\eqref{var_superdiff_2}, respectively, while
$L_{\alpha}^{0}\left(x \right)$ is a L\'evy symmetric $\alpha$-stable density.
This PDF is clearly self-similar with {\it space-time scaling}
$z=x/t^{\phi/2}$, being $p(x,t) = 1/t^{\phi/2} F(x / t^{\phi/2})$.

\noindent
The {\it formal} average of the fluctuation-dissipation relationship,
Eq. (\ref{cond_var}), is given by:
\be
\langle v^2 \rangle_{\rm eq} = \langle \tau \rangle \la D \ra\ .
\label{fluct-diss}
\ee
Being $\la D \ra=\infty$, this implies a physically meaningless infinite energy in the
equilibrium/stationary state: $\langle v^2 \rangle_{\rm eq} = \infty$ \footnote{
Actually, Eq. (\ref{fluct-diss}) is just a formal expression that, rigorously,
could not even be written when the mean diffusivity is infinite.
}.

\noindent
Considering that $D$ is connected with the mass $m$ and that, in real systems,
particles masses are finite, it is reasonable to assume that the PDF $f(D)$ a
maximum alllowed value for the diffusivity.
We then limit the possible values of the diffusivity $D$ by assuming a cut-off in the PDF
$f(D)$ at some maximum value $D_\mathrm{max}$\footnote{
A smoother (e.g., exponential) cut-off could be chosen, but we expect that
the particular choice of the cut-off does not substantially change the results.
}.
Consequently, the integral in Eq. (\ref{eq:space_fract}) becomes:
\be
\label{dmax_model}
p(x,t) = \int_0^{D_{\mathrm{max}}} {\cal G}\left(x,\sigma_X(t|D)\right) \frac{1}{D_*}
L_{\alpha/2}^{-\alpha/2} \left(\frac{D}{D_*} \right) \mathrm{d}D\ .
\ee
The PDF $p(x,t)$ is no more given by the symmetric L\'evy stable density
$L_\alpha^0$ as in Eq. (\ref{eq:space_fract}), but it still satisfies the
self-similarity condition: $p(x,t) = 1/t^{\phi/2} F(x/t^{\phi/2})$.

\noindent
The more interesting aspect is that model (\ref{dmax_model}) satisfies the
fluctuation-dissipation relationship averaged over $D$, Eq. (\ref{fluct-diss}),
thus also giving a finite energy.
This relationship can also be used to numerically estimate the value of 
$D_{\mathrm{max}}$ for given $\la D \ra$ or, equivalently, given
$\langle v^2 \rangle_{\rm eq}$ and $\langle \tau \rangle$:
\be
\langle D \rangle = \frac{{\langle v^2 \rangle}_{\rm eq}} {\langle \tau \rangle}
= \int_0^{D_{\mathrm{max}}} {D \, f(D) \, \mathrm{d}D}\ .
\label{dmax_evaluate}
\ee
Both this integral and the above integral in Eq. (\ref{dmax_model}) can be only
numerically evaluated, as a further analytical approach is not possible.

\noindent
For properly chosen values of $D_{\mathrm{max}}$, a power-law decay
$p(x,t) \sim 1/|x|^{1+\alpha}$ is expected to emerge in a intermediate range,
before a rapidly decaying cut-off appears at large $x$ values.
Further, all moments are finite, but they could have very large values
depending on the chosen value of $D_{\mathrm{max}}$. In any case, an anomalous
superdiffusive scaling is expected in the MSD:
$\langle x(t)^2 \rangle \sim t^\phi$.
%

\subsection{\label{sec:sim} Numerical simulations of the HEBP model}

\noindent
In order to verify the scaling features for the HEBP with truncated
diffusivity PDF, Eq. (\ref{dmax_model}), we carried out both Monte-Carlo (MC)
simulations of stochastic trajectories, computed from the Langevin equation
(\ref{eq:langevin_new}), and a direct numerical evaluation of the
integral in Eq. (\ref{dmax_model}).
In particular, the emergence of power-law behavior in the position PDF $p(x,t)$,
and the relative range of validity, needs to be numerically estimated.
%
%
%

\begin{figure}[!h]
\includegraphics[width=\linewidth]{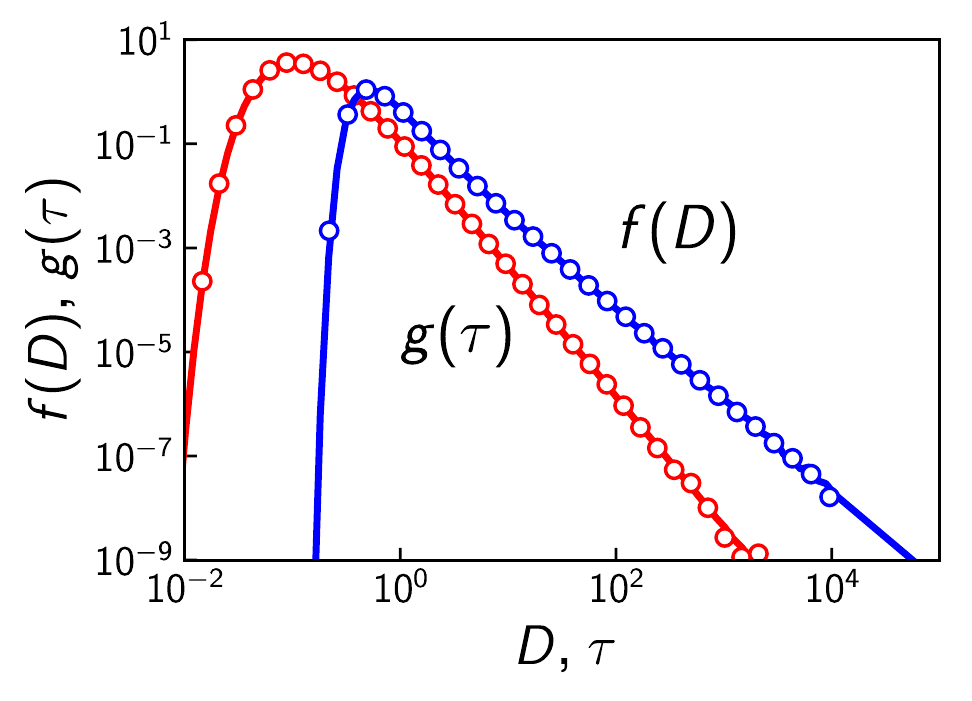}
\caption{
\label{fig:dstaus}(color online) Distributions $g(\tau)$
(red) and $f(D)$ (blue). Lines: theoretical expressions,
~\eqref{eq:tau_distr} and ~\eqref{eq:D_distr}.
Circles: histograms of the sample set of $\tau$ and $D$.
 $\tau_* = 1$; $D_* = 1$; $\alpha=3/2$; $\eta=1/2$.}
\end{figure}

\begin{figure}[!h]
\includegraphics[width=1\linewidth]{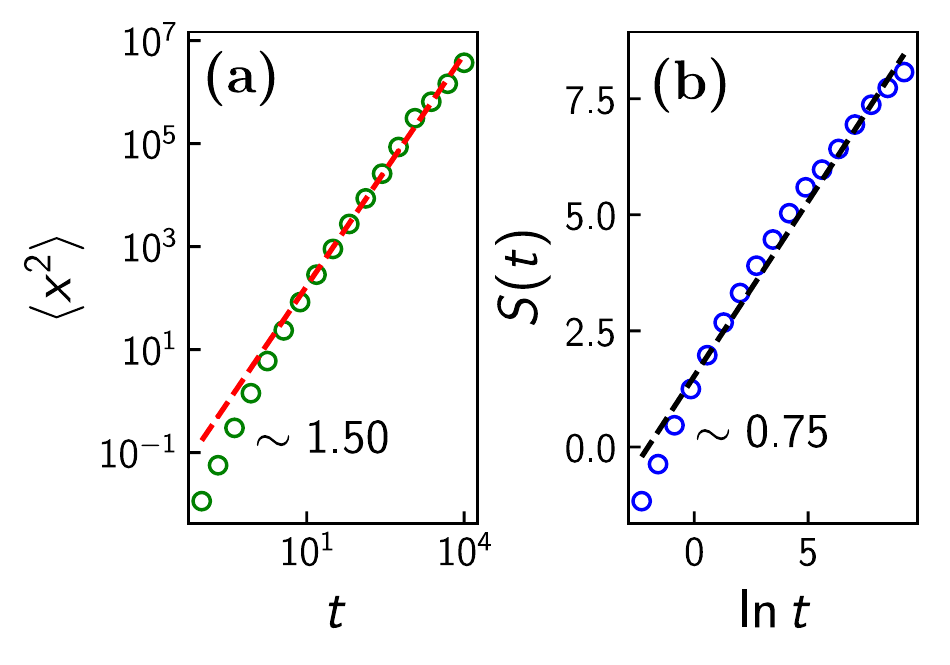}
\caption{
\label{fig:msd_shannon}(color online) \textbf{(a)} MSD computed from
the MC simulations (green circles) compared with the analytical prediction:
$\langle x^2(t) \rangle \propto t^\phi$; $\phi=3/2=1.5$ (red dashed line).
\textbf{(b)} Comparison of the DEA behavior computed from  MC simulations
(blue circles)
with DEA computed from the analytically obtained PDF,
Eq.~\eqref{dmax_model} (dashed black line). $\delta = 3/4 = 0.75$.
}
\end{figure}

\noindent
The numerical simulations of Langevin equation (\ref{eq:langevin_new}) and
Eq. ~\eqref{eq:kinematic} were carried out using the algorithms discussed in
\ref{num_algo}.
A sample set of couples $(\tau,D)$ was drawn from the distributions
$g(\tau)$ and $f(D)$, Eqs. (\ref{eq:tau_distr}-\ref{eq:D_distr}),
and stochastic trajectories were simulated.
In Fig. ~\ref{fig:dstaus} the theoretical distributions $g(\tau)$ and $f(D)$
are compared with the respective numerical histograms of drawn values of
$\tau$ and $D$.
In the numerical simulations the following parameters were used:
$\alpha=3/2$, $\eta=1/2$, $D_\mathrm{max}=10^4$,
initial conditions $x_{i,0} = 0$ and $v_{i,0} = 0$.
Being $\tau_*=1$, it results: $\la \tau \ra = \eta \tau_*/\Gamma(1/\eta)=1/2$.
From the numerical computation of Eq. ~\eqref{dmax_evaluate} we get:
$\la D \ra \simeq 8.31$ and, then:
$\la v^2 \ra_{\rm eq} = \la\tau \ra \la D \ra \simeq 4.16$.
%
Being $\phi=2-\eta=3/2$, we expect: 
$\langle x^2(t) \rangle \propto t^{3/2}$.
From Fig. ~\ref{fig:msd_shannon}(a) it is evident that the theoretical
scaling $\phi=3/2$ is numerically verified at sufficiently long times.

\noindent
In order to check the self-similarity of the position PDF:
\begin{equation}
\label{eq:pow_law_pdf_scale}
p(x, t) = \frac{1}{t^\delta} F \left( \frac{x}{t^\delta} \right),
\end{equation}
we use the Diffusion Entropy Analysis (DEA) \cite{grigolini_fractals01,allegrini_pre02,paradisi_csf15_pandora}, which is based on the computation of the Shannon
entropy of the diffusion process:
\begin{equation}
\label{eq:shannon}
S(t) = -\int_{-\infty}^\infty \mathrm{d}x\, p(x, t) \ln p(x, t) = A +
\delta \ln t\ .
\end{equation}
$\delta$ is the space-time scaling of the PDF.
For monoscaling diffusion \footnote{
  Monoscaling diffusion processes belong to the general class of
  monoscaling/monofractal processes or signals, defined by the
  condition: $X(at) = a^H X(t)$.
},
Eq. ~\eqref{eq:pow_law_pdf_scale} holds exactly, so that the PDF
$p(x,t)$ is self-similar with self-similarity index $\delta$, thus also equal
to the Hurst exponent $H$.
The theoretical expectation for the HEBP is:
$\delta = H = \phi/2 = 1-\eta/2$.
%
%
%
The DEA was computed using the histograms estimated from numerical MC
simulations and from the numerical computation of the analytical expression
(\ref{dmax_model}).
The comparison of the two different estimates is shown in
Fig. ~\ref{fig:msd_shannon}(b). It is evident that the DEA computed from the
analytical expression shows very good agreement with the theoretical scaling:
$\delta=\phi/2 = 1-\eta/2=3/4$.
On the contrary, in the DEA computed from MC simulations a net straight line
in the graph $(\ln t,S(t))$ does not emerge clearly in the studied range,
even if a rough agreement with the theoretical expectation is seen.
This is probably due to statistical limitations of MC simulations, thus
proving that estimation of scaling in such processes could be quite a delicate
task when dealing with real experimental data.
%

\begin{figure}[!h]
\includegraphics[width=1\linewidth]{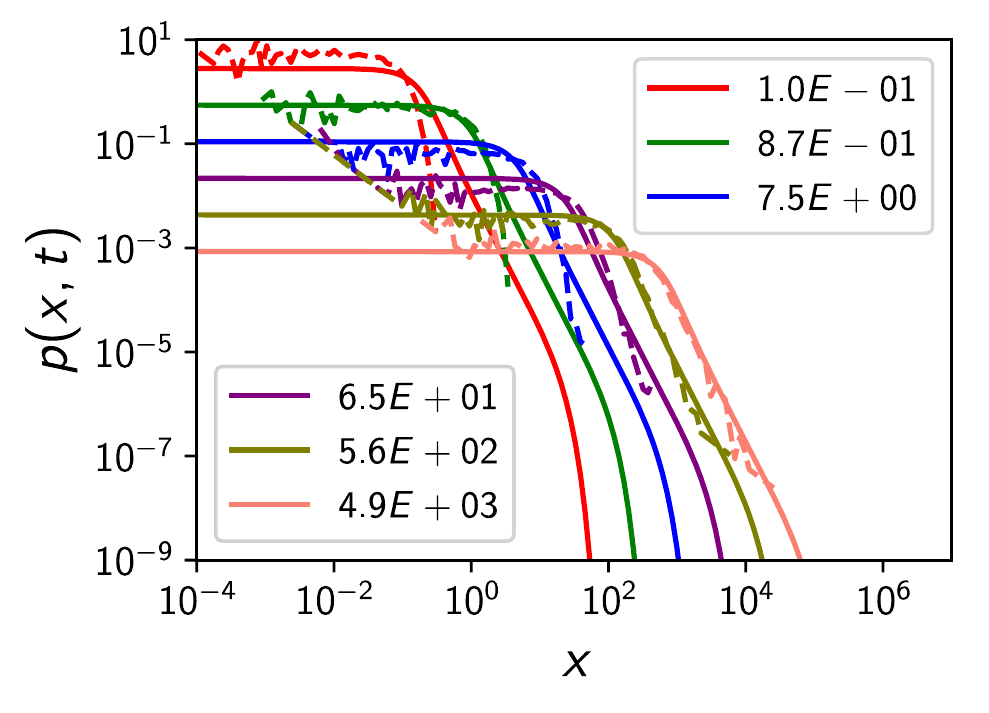}
\caption{\label{fig:pdf9600_cmp_an}(color online) Comparison of the
  coordinate PDFs of the MC simulated motion (dashed lines) with that
  obtained from the analytical expression ~\eqref{dmax_model} (straight lines)
  for different times.}
\end{figure}

\noindent
In Fig.~\ref{fig:pdf9600_cmp_an} we compare the coordinate PDFs computed from
Eq.~\eqref{dmax_model} with those evaluated from the MC simulations (being
diffusion symmetrical, the PDFs are plotted in the range $x>0$).
The analytical expression clearly shows a well-defined power-law tail in
a intermediate range of $|x|$: $p(x, t) \propto 1/|x|^{1+\alpha}$,
followed by a rapid cut-off for large $|x|$.
Regarding the space-time scaling $z=x/t^{\delta}$, in agreement with DEA, the
analytical PDFs have an exact self-similarity index $\delta=3/4$.
Conversely, the decay of PDFs derived from MC simulations is slightly more
complicated, but the general behavior is compatible with the analytical one.
Even the self-similarity space-time scaling roughly approximates the theoretical
expectation $\delta=3/4$, but small deviations are evident, is agreement with
the DEA displayed in Fig. ~\ref{fig:msd_shannon}(b) (blue circles).

\subsection{Heterogeneous ensemble of Brownian particles and superstatistics}
\label{superstat}

\noindent
Superstatistics approach takes into account large-deviations of intensive
quantities of systems in nonequilibrium stationary states
\cite{beck_prl01,beck_pa03,abe_etal-pre-2007} 
and it was motivated by some preliminary success obtained when fluctuations of 
parameters were considered \cite{wilk_etal-prl-2000,beck-epl-2002}.
In general, superstatistics is successful to model:
turbulent dispersion considering energy dissipation fluctuations
\cite{beck_prl01,reynolds-prl-2003},
renewal critical events in intermittent systems \cite{paradisi_cejp09,akin_jsmte09},
and for different distributions of the fluctuating intensive quantities
different effective statistical mechanics can be derived
\cite{beck_pa03}, e.g., Tsallis statistics with $\chi^2$-distribution.

\noindent
The main idea of superstatistics is that a Brownian test particle
experiences fluctuations of some intensive parameters by moving from cell to cell
\cite{beck_pa03}.  Following this idea, the random value of
the fluctuating parameter is generated at any change of cell.
The main assumption behind
this picture is that each cell is in equilibrium during the residence time of
the particle: within the cell there are no fluctuations but a different
value assigned to each cell. The local
value of the fluctuating parameter changes in the
various cells on a time scale that is much longer
than the relaxation time that the single cells need
to reach local equilibrium.
This means that the fluctuating parameter
follows a slow dynamics and then the integration over the fast
variable is taken after the integration over the slow variable which is in
opposition to what an adiabatic scheme requires \cite{abe_fp14}.
This fact can be considered just an order of integration that does not
affect the computation of the expected values but it is more deep when
the entropy is considered \cite{abe_etal-pre-2007,abe_fp14}.  This
inconsistency is solved by considering a dynamical equation also for
the slow fluctuating quantity \cite{abe_fp14}, an example of such
dynamical equation was already considered in Ref.
\cite{reynolds-prl-2003}.

\noindent
The HEBP approach is clearly based on a different picture, even if
the superposition of Langevin equations may suggest some analogies.
Here the superposition gives rise to anomalous diffusion because it reproduces the
effects of the ensemble heterogeneity.
In fact, in the present approach the fluctuations are not due to different values in
different cells but to the population of density (mass divided by volume)
of the ensemble.
As a consequence of this, the present approach does not take into account slow and fast
dynamics and then the issue concerning the order of integration does not arise.

\subsection{Anomalous-to-normal transition}
\label{anom-to-normal}

\noindent
Here we briefly show the effect of limited statistics of $\tau$ on the diffusion scaling.
Statistical limitation in the number of $\tau$ randomly drawn from $g(\tau)$ 
also results in the existence of a maximum relaxation time $\tau_{\rm max}$.
Thus, the statistical limitation in the sample set of $\tau$ mimicks the existence
of a $\tau_{\rm max}$ in real experimental systems.
This is a reasonable assumption, also considering the relation of
$\tau$ with mass and size of the particle (see previous Section \ref{model_1}), whose
distributions are necessarily limited.
%
\begin{figure}[!h]
\includegraphics[width=\linewidth]{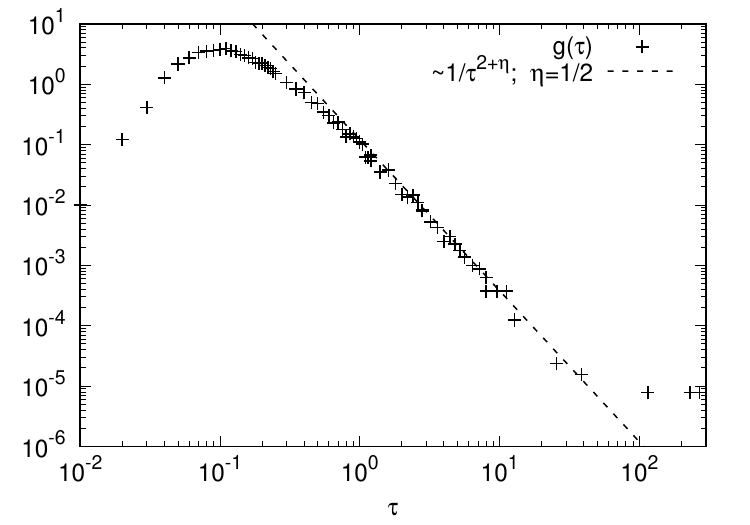}
\caption{
 \label{fig:tau-histo}(color online) Histogram of a sample set of $\tau$ limited to
 $10000$ draws from $g(\tau)$. $\tau_* = 1$.
}
\end{figure}
In Fig. \ref{fig:tau-histo} we report the histogram for a sample set with $10000$
random draws from $g(\tau)$. The parameters are: $\eta=1/2$, $\la\tau\ra=1/2$.
It can be seen that, for this limited statistics, $g(\tau)$ is well-reproduced up to a
value of
$\tau$ less than $10$, while for larger values there are fluctuations and, for $\tau$ greater
than about $30-50$, also some apparent outliers are seen till a maximum value
$\tau_{\rm max} = 297.2$.
The experimental/numerical mean relaxation time is $\la\tau\ra_{\rm exp}=0.52$.
\begin{figure}[!h]
  \includegraphics[width=\linewidth]{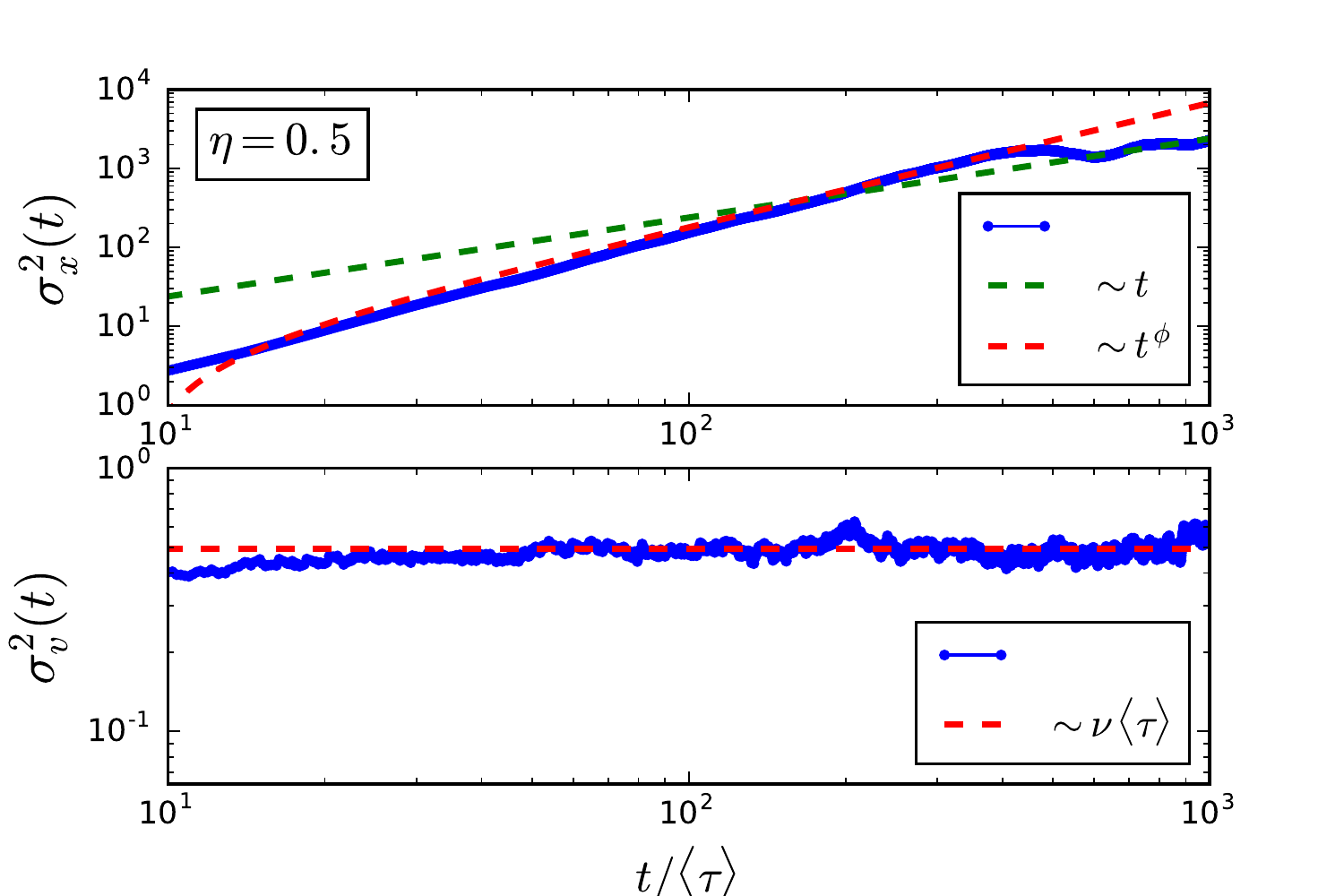}
\caption{
  \label{fig:anom-norm}(color online) Numerical simulations of the heterogeneous
  ensemble of Brownian particles for the $\tau$ sample set of Fig. \ref{fig:tau-histo}.
Top panel: position MSD; bottom panel: velocity MSD.
}
\end{figure}

\noindent
The maximum relaxation time $\tau_{\rm max}$ can be considered a kind of time scale after which all trajectories reach
the condition of a variance increasing linearly in time, even if with different multiplicative
factors. As a consequence, we expect normal diffusion to occur in a very long time
regime.
This is confirmed in Fig. \ref{fig:anom-norm}, where the anomalous diffusive scaling
$\phi=2-\eta=3/2$ emerges in the approximate time interval
$[15 \la \tau \ra,450 \la \tau \ra]$, after which there is a transition to a normal
diffusion regime starting at $t \sim 600 \la \tau \ra$.
It is worth noting that, in general, the transition time scale does not depend only on
$\tau_{\rm max}$, but also on the detailed statistics of the numerical histogram in the
neighborhood of $\tau_{\rm max}$ itself. In particular, a situation where $\tau_{\rm max}$ is
an outlier is quite different from a condition where $\tau$-set is, in some sense, dense near
$\tau_{\rm max}$, which cannot then be considered an outlier.

%
%

\noindent
In summary, we can argue that, depending on the experimental/numerical set of relaxation
times $\tau$, our HEBP model reproduces a transition from anomalous to normal
diffusion.

\section{\label{sec:lw}Comparison with L\'evy walk models}

LW is one of the best known models of anomalous diffusion with finite
MSD and was firstly introduced by Shlesinger, Klafter and Wong
in 1982 \cite{shlesinger_random_1982}.
The number of papers devoted to LWs is very large (see, e.g., \cite{allegrini_pre02,klages_anomalous_2008,paradisi_csf15_pandora,magdziarz_explicit_2016,taylor-king_fractional_2016,dybiec_pre17,aghion_epjb18}) and
a quite recent and complete review can be found in Zaburdaev et al., 2015
\cite{zaburdaev_rmp15}.
LWs have been applied to many phenomena, but surely the most promising
and widespread applications are in the modeling of search strategies,
such as bacteria foraging through run-and-tumble motion
\cite{viswanathan_2011,ariel2015swarming,zaburdaev_rmp15}.
%
%
%
Unlike L\'evy flights, where the particle is allowed to make large
jumps in a whatever short time step (theoretically zero in the time-continuous
limit), thus giving instantaneous infinite velocities and discontinuous paths,
in LW models the particle moves with a finite speed. Such speed remains
constant throughout a random duration time, also called Waiting Time (WT).
After this WT, velocity randomly changes according to an assigned walking rule
and it remains constant for another random WT.
Thus, even if there are events with discontinuous acceleration, LWs have
continuous velocity.
When WT have a constant value, equivalent to a fixed time step, and velocity
MSD is finite, LW reduces to a standard Random Walk with ballistic diffusion:
$\la x^2\ra \sim t^2$. For WTs with finite mean, ballistic diffusion also
occurs, but in the long-time limit (e.g., exponentially distributed WTs).
Interestingly, the LW also displays {\it strong anomalous diffusion}, also
known as {\it multiscaling/multifractal} diffusion
\cite{ott2002,kantelhardt_pa02}.
Multiscaling detection algorithms are usually based on the analysis of
fractional moments:
\begin{equation}
\label{eq:frac_moments}
\langle |x|^q \rangle = \int_{-\infty}^{\infty} \mathrm{d}x |x|^q p(x,
t) = M_q \cdot t^{\lambda(q)}\ ,
\end{equation}
where $\lambda(q) = q H(q)$, being $H(2)$ the well-known Hurst exponent or
second-moment scaling.
A complex system is multiscaling when $H$ changes with the moment order $q$
and the particular multiscaling features are defined by the behavior of
the function $H(q)$. Conversely, a constant $H$, thus independent of $q$,
is associated with monoscaling systems: $\langle |x|^q \rangle \sim t^{q H}$.

\noindent
Here we consider two different LW models that differ for the
velocity distribution. The first one is the most classical one with randomly
alternating velocities, i.e., constant speed $|V_{_{\rm LW}}|$ and randomly
changing direction according to a coin tossing prescription
\cite{shlesinger_random_1982,allegrini_pre02,zaburdaev_rmp15}.
We limit here to the case $V_{_{\rm LW}} = \pm 1$.
In the second one, we consider a continuous and symmetric random variable for
the velocity \cite{zaburdaev_rmp15}.
In both LW models the velocity is constant throughout a WT of duration
$\delta t_i = t_{i+1}-t_i$ and randomly changes
in correspondence of the critical event $i+1$, whose occurrence time $t_{i+1}$
marks the passage from the WT $\delta t_i$ to the next WT $\delta t_{i+1}$.
For the WT distribution, we consider the following PDF \cite{paradisi_romp12}:
\begin{equation}
\label{eq:lw_time_distr}
\psi(\delta t) = \frac{\left(\mu-1\right)T^{\mu-1}}
    {\left( T + \delta t \right)^{\mu}},
\end{equation}
where $\mu > 1$ and $T$ is a reference time scale. The power-law tail emerges
in the range $\delta t \gg T$. In the following we set $T=1$.
The superdiffusive sub-ballistic behavior (the
one we are interested in) is revealed when $2<\mu<3$.
In the LW with alternating velocities, this regime is characterized by a
central part of the PDF $p(x,t)$ that is well approximated by a symmetric
L\'evy stable density $L_\alpha^0$ with stability index
$\alpha = 1/(\mu-1)$ \cite{allegrini_pre02,paradisi_romp12}.
At sufficiently large $|x|$, the PDF is abruptly truncated by ballistic peaks located at
$x = \pm V_{_{\rm LW}} t$, which corresponds to the ballistic motion of paths
whose first WT is longer than $t$.
\begin{figure}[!h]
\centering \includegraphics[width=1\linewidth]{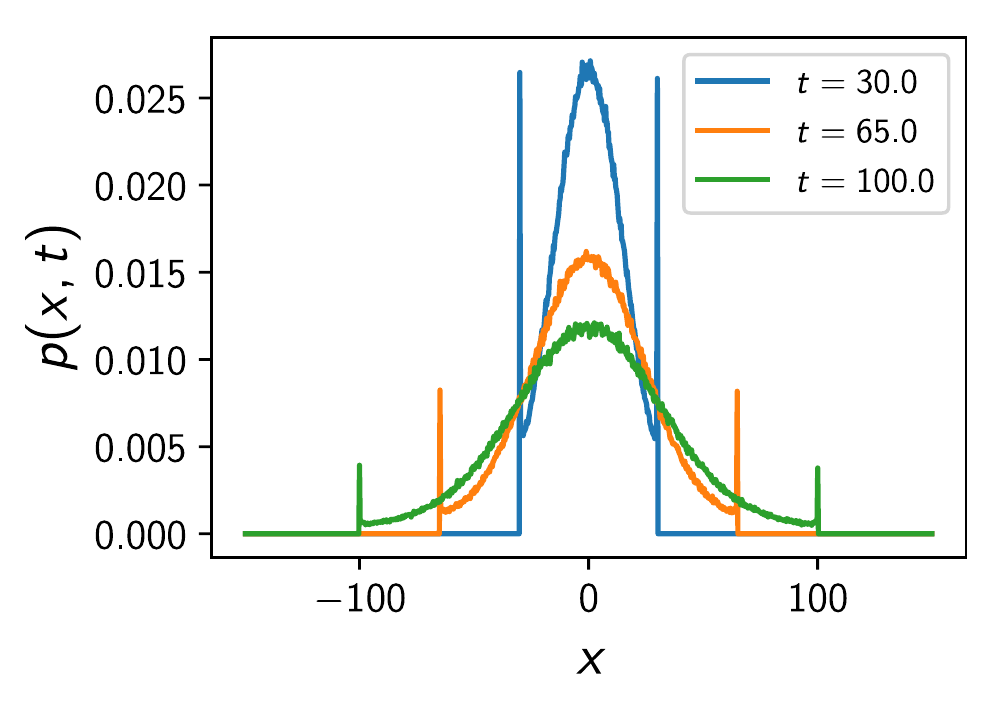}
\caption{\label{fig:lw_pdfs_lin}L\'evy walk probability density
  functions: the central part looks like L\'evy $\alpha$-stable law
  cut by ballistic fronts at $x = \pm V_{_{\rm LW}} t$}.
\end{figure}
In Fig.~\ref{fig:lw_pdfs_lin} the PDFs at different times computed from a MC
simulation of LW with alternating velocities are reported. The ballistic
peaks truncating the PDFs are evident.

\noindent
Similarly to HEBP, Eq. ~\eqref{dmax_model}, this LW model displays a power-law
decay in a intermediate range of $|x|$ followed by an abrupt cut-off,
thus resulting in the finiteness of moments and, in particular, of the MSD:
$\langle x^2 \rangle \propto t^{4-\mu}$
\cite{allegrini_pre02,zaburdaev_rmp15}.
For the LW with randomly alternating velocities and WT-PDF given by
~\eqref{eq:lw_time_distr} \footnote{
This is valid for all WT-PDFs with fat tails: $\psi(\tau) \sim 1/\tau^\mu$.
},
the fractional moments are given by \cite{zaburdaev_rmp15}:
\be
\lambda(q) = \left\{
\begin{array}{l}
  q/(\mu-1)\ ; \quad \quad\  \ q \le \mu-1\ ;\\
  \ \\
  q - (\mu-2)\ ; \qquad q > \mu-1\ .
\end{array}
\right.
\label{levy_multiscaling}
\ee
It is then easily seen from this formula that the LW with randomly alternating
velocities obeys a biscaling law, with a given scaling for low-order moments
and another one for high-order moments.

\noindent
In the following we compare four different cases: two LW models
(with alternating and continuous velocities, respectively) and our HEBP model for
two different sets of parameters chosen to fit the LW models.
In particular:
\begin{itemize}
\item[(i)]
  L\'evy walk with randomly alternating velocity rule. WT-PDF given by
  Eq. ~\eqref{eq:lw_time_distr} with $\mu=5/2$. Coin tossing prescription for
  the change of direction. We set $V_{_{\rm LW}} = \pm 1$, so that
  $\langle v^2 \rangle_{\rm eq} = \la v^2_{_{\rm LW}} \ra = 1$;
\item[(ii)]
  HEBP analytical model, Eq. (\ref{dmax_model}), with parameters: $\alpha=3/2$,
  $\eta = 1/2$,
  $\langle v^2 \rangle_{\rm eq} =  \langle \tau \rangle \langle D \rangle = 1$.
  It results: $\la \tau \ra = 1/2$, $\la D \ra = 2$, $\phi=2-\eta=3/2$ and
  $D_\mathrm{max} = 40.1$
  (numerically calculated from Eq. ~\eqref{dmax_evaluate});
\item[(iii)]
  L\'evy walk with random continuous velocity. WT-PDF given by Eq.
  ~\eqref{eq:lw_time_distr} with $\mu=5/2$. The velocity PDF is symmetric and evaluated
  from the stationary state of MC simulations carried out for the HEBP,
  Eq. ~\eqref{dmax_model}.
  MC simulation parameters are the same as in the next case (iv).
  The random generation of velocities was performed with the inverse
  transform sampling method;
\item[(iv)]
  HEBP analytical model, Eq. (\ref{dmax_model}), with parameters:
  $\alpha=3/2$,  $\eta = 1/2$,
  $\langle v^2 \rangle_{\rm eq} = \langle \tau \rangle \langle D \rangle = 8.127$.
  It results: $\la \tau \ra = 1/2$, $\phi=2-\eta=3/2$, $\la D \ra = 16.254$ and
  $D_\mathrm{max} = 1.43\cdot 10^5$
  (numerically calculated from Eq. ~\eqref{dmax_evaluate});
\end{itemize}
The PDFs $p(x,t)$ of the HEBP, cases (ii)  and (iv), are obtained, for
different times, by means of numerical evaluation of Eq. ~\eqref{dmax_model}.
Then, DEA $S(t)$ and MSD $\la x^2(t) \ra$ are computed by the calculated PDFs.
Conversely, the paths of LW models (i) and (iii) are computed by means of MC
stochastic simulations and, then, PDFs, DEA and MSDs are evaluated by
statistical analysis of the sample paths.

\begin{figure}[tbp]
  \centering
  \includegraphics[width=\linewidth,height=0.6\linewidth]{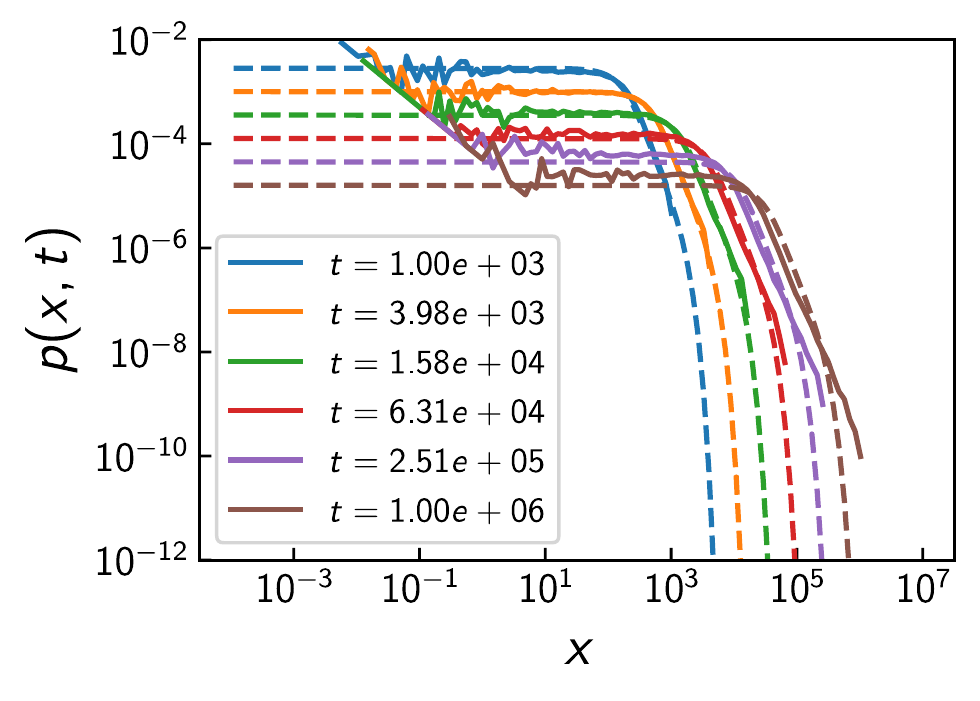}\\
  \includegraphics[width=\linewidth,height=0.6\linewidth]{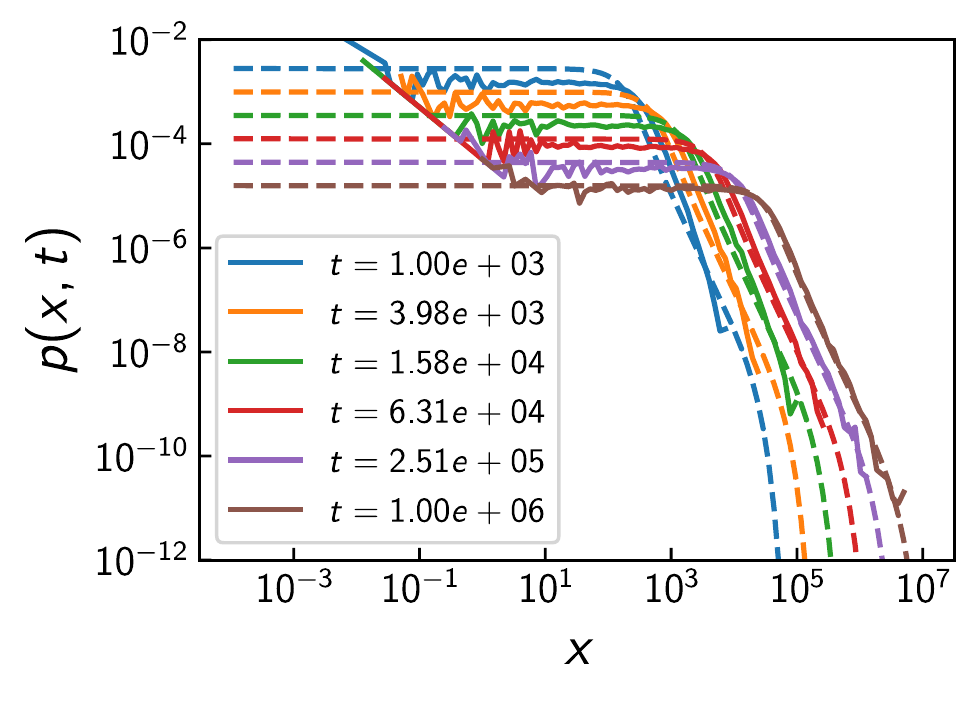}
  \caption{\label{fig:lw_an_1}
    Comparison of PDFs $p(x,t)$ for the four model cases (i-iv):
    {\bf Top panel}: LW model (i) with randomly alternating velocities
    (solid lines) fitted by HEBP, case (ii) (dashed lines);
    {\bf Bottom panel}: LW model (iii) with continuous random velocities
    (solid lines) fitted by HEBP, case (iv) (dashed lines). 
  }
\end{figure}
\begin{figure}[tbp]
  \centering
  \includegraphics[width=\linewidth,height=0.6\linewidth]{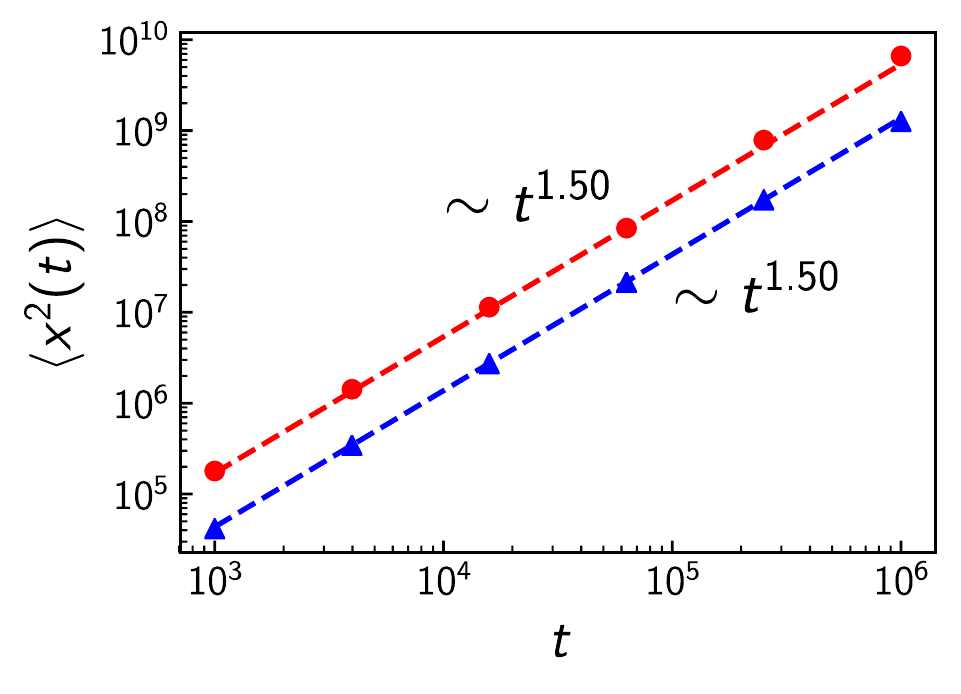}\\
  \includegraphics[width=\linewidth,height=0.6\linewidth]{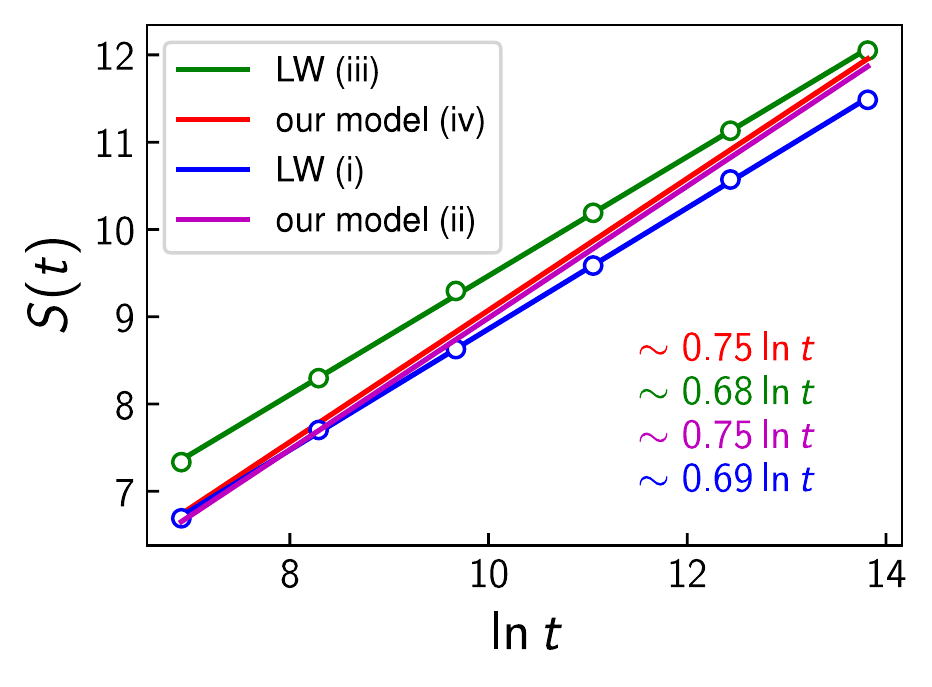}
  \caption{\label{fig:lw_an_2}
    Comparison of MSD $\la x^2 \ra(t)$ and DEA $S(t)$, Eq.~\eqref{eq:shannon}),
    for the four model cases
    (i-iv):    
    \textbf{Top panel}:
    LW model (i) with randomly alternating velocities (blue triangles);
    HEBP, case (ii) (dashed blue line);
    LW model (iii) with random continuous velocities (red circles);   
    HEBP, case (iv) (dashed red line);
    \textbf{Bottom panel}:
    LW model (i) with randomly alternating velocity (blue dots and line);
    HEBP, case (ii) (purple line);
    LW model (iii) with random continuous velocity (green dots and line);
    HEBP, case (iv) (red line).
  }
\end{figure}
\begin{figure}[tbp]
  \centering
  \includegraphics[width=\linewidth,height=0.6\linewidth]{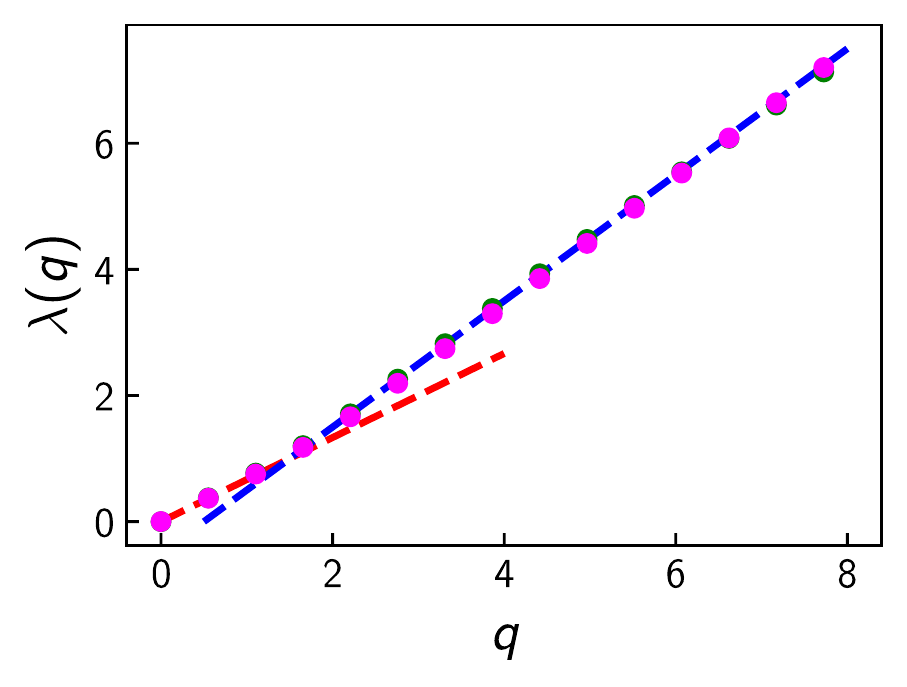}\\
  \includegraphics[width=\linewidth,height=0.6\linewidth]{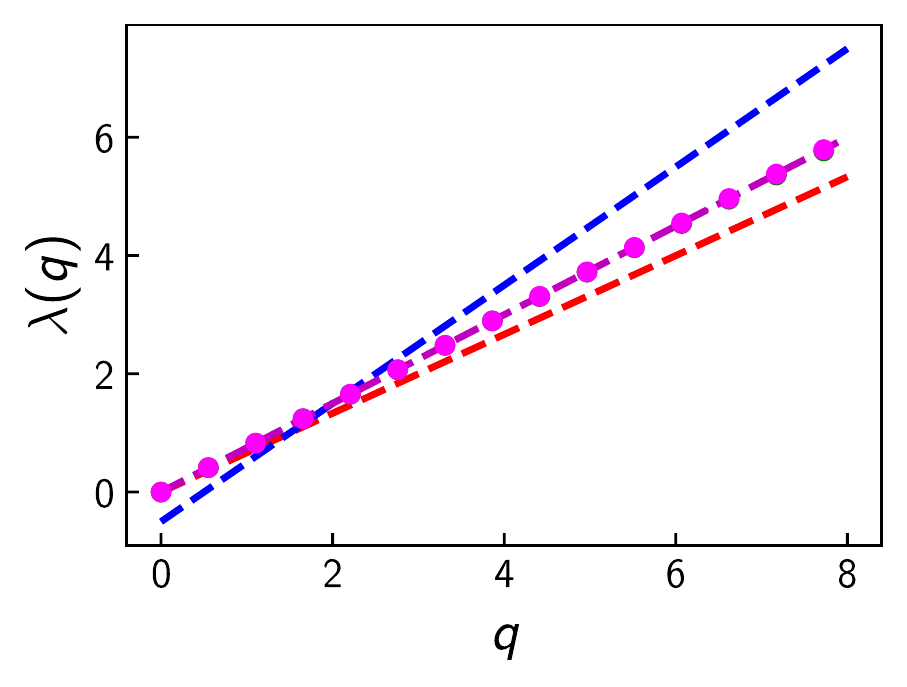}
  \caption{\label{fig:lw_an_3}  
  \textbf{Top panel}:
  fractional moments, (Eq.~\eqref{eq:frac_moments}), of LW models (i) and (iii)
  (purple and green points, respectively -- almost coincide),
  dashed lines are analytical asymptotes given by
  Eq. ~\eqref{levy_multiscaling};
  \textbf{Bottom panel}:
  fractional moments of HEBP, cases (ii) and (iv) (purple and green points,
  respectively -- almost coincide), dashed lines are the same analytical
  asymptotes as in Top panel.
  }
\end{figure}

\noindent
The results are gathered in Figs. ~\ref{fig:lw_an_1},~\ref{fig:lw_an_2} and
~\ref{fig:lw_an_3}.
In Fig. ~\ref{fig:lw_an_1} the comparisons of position PDFs $p(x,t)$ for
different times are shown: LW model (i) fitted by HEBP, case (ii) (top panel)
and LW model (iii) fitted by HEBP, case (iv).
It is evident that the two LW models (alternating/continuous velocities) have
very similar behaviors and that, for both models, it is always possible to
find a parameter set for the HEBP to be comparable with LW models.
In particular, HEBP well reproduces the power-law decay in the intermediate
range, while for $x$ greater than the ballistic peaks of LW models
an exponential cut-off emerges in the PDFs of HEBP.
However, the space-time scaling $\delta$ is clearly different for the two
models,  as it is clear from the slight shifts between solid and dashed lines
in both panels.
Then, for fixed set of parameters, the quality of the fit is not the
same for all PDF times.
In particular, in the top panel (models (i) and (ii)) the best fit
is made at time $t=10^3$, so that the less accurate agreement is at the
longest time $t=10^6$. On the contrary, in the bottom panel the best fit is
made at $t=10^6$ and, consequently, the worst agreement is at the first
displayed time $t=10^3$.

\noindent
In Fig. ~\ref{fig:lw_an_2} we report the MSD $\la x^2 \ra$ (Top panel) and
the DEA $S(t)$ (Bottom panel) for the four model cases.
Numerical simulations and calculations of all models (i-iv)
(LW models and HEBP) reproduce the expected power-law dependence
$\la x^2 \ra \sim t^{3/2}$ with a very good agreement.
The DEA $S(t)$ shows the net differences of the self-similarity
index $\delta$ among LW and HEBP, thus making it more evident the
different space-time scaling seen in Fig. ~\ref{fig:lw_an_1}.
The numerical scaling is also in agreement with theoretical values:
$\delta=1/(\mu-1)=2/3 \simeq 0.67$ for LWs and $\delta=1-\eta/2 = 3/4$
for HEBP.

\noindent
Finally, in order to explore the multiscaling character of the models,
Fig. ~\ref{fig:lw_an_3} show the results for the evaluation of fractional
moments.
In top panel the fractional moments of LW model (i) and (iii) are compared,
while bottom panel compares HEBP, cases (i) and (iv).
The LW models (i) and (iii) have exactly the same behavior for $\lambda(q)$,
thus in agreement with the expected multiscaling and, in particular, the
biscaling law of Eq. ~\eqref{levy_multiscaling}.
The behavior of our numerical simulations of HEBP, cases (ii) and (iv), is
also found to be very similar to each other and in agreement with theoretical
predictions, that is, it shows a well-defined monoscaling.
The space-time scaling is the same for both parameter sets, as it
depends only on $\eta$, being $\la X^q \ra \sim t^{\lambda(q)}=t^{qH(q)}$ with:
\be
H(q)=\phi/2=1-\eta/2,
\label{our_scaling}
\ee
and, for HEBP, this is also equal to the self-similarity index:
$\delta=H(q)=\phi/2$ (see Eq. ~\eqref{eq:space_fract}).

\section{\label{discuss}Discussion and concluding remarks}

\noindent
We here introduced and discussed a model based on the idea of a standard
friction-diffusion process in a strongly heterogeneous condition
with inverse power-law distributions of parameter's populations
(Eqs. ~\eqref{eq:tau_distr} and ~\eqref{eq:D_distr}).
%
%
We considered the Langevin equation for an Ornstein--Uhlenbeck process with
randomly distributed relaxation/correlation times $\tau$ and
diffusivities $D$.
The model with random $\tau$ population and constant diffusivity gives
a Gaussian process with long-range correlation and anomalous diffusion scaling.
The moments $\la |x|^q \ra$ are finite, as well as the energy
$\la v^2 \ra_{\rm eq}$.
However, anomalous transport often displays power-law decays in the position
PDF $p(x,t)$ that cannot be reproduced by Gaussian processes, even if
long-range.
In order to extend this model to non-Gaussian PDFs with power-law tails,
a random $D$ population is needed. This is obtained by means of the distribution
$f(D)$ given in Eq. ~\eqref{eq:D_distr}. However, this distribution has an infinite
mean and determine infinite moments for the velocity PDF and, thus, infinite
energy. This is an unphysical condition, which also prevents to get a
fluctuation-dissipation relation.
For this reason we adopted a more realistic assumption by imposing a
cut-off maximum value $D_{\rm max}$ for the diffusivity population, from
which Eq. ~\eqref{dmax_model} follows.

\noindent
We proved that, similarly to LWs, our proposed HEBP model can take into account
intermediate power-law decays in the PDF.
%
%
%
Unlike LWs, where power-law
is truncated by the ballistic peaks due to the underlying WT statistics,
in HEBP the power-law is truncated by an exponential cut-off in the regime of
large $x$. In experimental data, this is often associated with lack of
statistics or presence of noise
\cite{allegrini_pre10,paradisi_csf15_pandora,paradisi_springer2017}, but it is
also recognized to be reminiscent of heterogeneous media \cite{lanoiselee_jpamt18}.

\noindent
In summary, we derived a model in a physical framework involving
{\bf heterogeneity} and, then, a {\bf population} of parameters characterized
by given inverse power-law distributions.
Our model follows from a superposition of standard Gaussian processes
with stationary and independent increments.
This model therefore has the following properties:
\begin{itemize}
  \vspace{-.2cm}
\item[(1)] {\bf long-range} correlations $R(t)\sim1/t^\eta$ and anomalous {\bf superdiffusive} scaling in the variance: $\la x^2 \ra \sim t^\phi$ ($\phi=2-\eta$; $1\le\phi<2$);
  \vspace{-.2cm}
\item[(2)] {\bf finite moments} $\la |x|^q\ra$, {\bf finite energy}
  $\la v^2\ra_{\rm eq}$ and a fluctuation-dissipation relation:\\
  $\la v^2\ra_{\rm eq} = \la \tau \ra \la D \ra$;  
  \vspace{-.2cm}
\item[(3)] both an intermediate range with {\bf power-law decay} $1/|x|^{1+\alpha}$
  and an asymptotic range with exponential cut-off in the PDF $p(x,t)$;
  \vspace{-.2cm}
\item[(4)] space-time {\bf monoscaling} behavior: $x \sim t^\delta$ \\
  ($\delta=\phi/2$);
\vspace{-.2cm}
\item[(5)]
  $\alpha$ and $\delta$ are independent scaling parameters;
  \vspace{-.2cm}
\item[(6)] a {\bf transition} from anomalous (intermediate time regime) to normal
  diffusion (long time regime).
\end{itemize}
This last point implies that the (mono-)scaling index $\delta$ is a function
of time, being $\delta(t) \ne 1/2$ for $t$ much less than the maximum relaxation time
$\tau_{\rm max}$, and $\delta(t) = 1/2$ for $t\gg \tau_{\rm max}$.

\noindent
Properties (1-3) are similar to those displayed by LW models, apart
from the exponential cut-off of HEBP, which could be difficult to
distinguish from the cut-off in the LW-PDF when dealing with experimental
data.

\noindent
On the contrary, property (4) is not satisfied by LWs, which obey
the {\bf biscaling} law ~\eqref{levy_multiscaling},
distinctly different from the {\bf monoscaling} behavior of HEBP.
Also the crucial property (5) is not seen in LWs.

\noindent
Further, L\'evy Walks do not reproduce the right space-time scaling $\delta$
of HEBP neither in the PDF's central part, which is that part of the PDF
more similar to a pure L\'evy stable density.
In fact, in our HEBP model the space-time scaling $\delta$ and the
power-law decay of the probability distribution $\alpha$ are {\bf independent
parameters}, while they are not in LWs.
LW is only driven by the parameter $\mu$ associated with the underlying
trapping mechanism described by Eq. ~\eqref{eq:lw_time_distr}.
The additional assumption of jumps coupled with WTs triggers the emergence of
anomalous superdiffusion, thus directly affecting the space-time self-similarity
index $\delta_{_{\rm LW}}$\footnote{
It is interesting to note that this is also true when the velocity PDF,
even if characterized by a power-law decay, has finite variance, and finite
higher-order moments, due to the cut-off (always present in real experimental
data).
}.

Thus, when the LW-PDF is characterized by the decay: $p(x,t)\sim 1/|x|^{\mu}$,
the scaling $\delta_{_{\rm LW}}$ is constrained, by the jump-WT coupling,
to obey the relationship \cite{allegrini_pre02, zaburdaev_rmp15}:
\be
\delta_{_{\rm LW}}=\frac{1}{\mu-1}=\frac{1}{\alpha_{_{\rm LW}}}\ .
\label{lw_delta}
\ee
where $\alpha_{_{\rm LW}} = \mu-1$ represents the L\'evy stability index.
As known, this is well-established in the intermediate range where the LW-PDF
is more similar to a pure L\'evy stable density $L^0_{\alpha_{_{\rm LW}}}$.
It is important to notice that the above relationship among $\delta$ and
$\alpha$ can be also satisfied by our HEBP model for particular parameter choices,
i.e., given the experimental $\alpha$, for $\eta=2-2/\alpha$.

\noindent
Another important aspect worthy of discussion is the physical basis of the
considered models.
HEBP models directly follow from an heterogeneity assumption applied to a
standard Gaussian process, whether the origin of heterogeneity is (in the
medium or in the particle parameters).
Thus, HEBP models, which are based on the same idea of the ggBM
\cite{mura-phd-2008,pagnini_etal-ijsa-2012}, are derived from a
physical background directly involving the idea of a complex heterogeneity
and indeed we expect HEBP models to be
more suitable to heterogeneous transport phenomena.
Conversely, LWs should better fit phenomena where trapping plays a fundamental role.
%
%
%

\noindent
As already said, many authors have recently been focusing on position transport
models with heterogeneous diffusivities, e.g., DDMs
\cite{chubynsky_etal-prl-2014,chechkin_prx17,jain_jcs17,lanoiselee_jpamt18,sposini_njp18}.
%
%
In some sense, HEBP models belong to the class of transport models with random
diffusivity, i.e., HDMs. However, unlike other ones, the HEBP model here discussed explicitly describes the velocity
dynamics, thus including the often neglected but crucial role of viscous
relaxation time $\tau$. 
More precisely, we here refer to the relaxation time of
the velocity, a physical parameter whose relationship with medium/fluid
properties is well-established.
As seen above, the role of heterogeneity in the relaxation time $\tau$
is taken into account and modeled through a
population with inverse power-law distribution $g(\tau)$.
Another important aspect is that the here proposed HEBP model is derived from a
standard friction-diffusion process having finite energy and satisfies a
fluctuation-dissipation theorem.

\noindent
From the above discussion, we can finally suggest a possible statistical
recipe to distinguish the best modeling approach among LWs and HEBP models
starting from a set of experimental transport data. This is of great interest
when the underlying mechanism, heterogeneity or trapping, driving anomalous
diffusion is not yet clear.
If the experimental PDF displays a power-law decay, it is possible to
apply a best fit procedure to get $p(x,t)\sim 1/|x|^{1+\alpha_{_{\rm exp}}}$ for some
$\alpha_{_{\rm exp}}<2$, where 'exp' stands for {\it experimental}.
Then, fractional moments and the function  $\lambda(q) = q H(q)$
can be computed. DEA can be applied to compute the self-similarity index
and let us assume that a well-defined $\delta$ exists.
Then, we have the indices $\alpha_{_{\rm exp}}$, $\delta$ and $\lambda(q)$.
If the data are monoscaling, then HEBP could be a good candidate and
we have two parameters to independently fit the PDF scaling ($\alpha$) and the
moment scaling ($H(q)=\phi/2=$ constant).
The exponential cut-off could be another clue towards HEBP \cite{lanoiselee_jpamt18},
but actually an exponential cut-off is usually seen in the tail of experimental PDFs due to
lack of statistics and/or presence of instrumental/environmental noise.
If data are multiscaling, then there are two possibilities: (i) biscaling
law ~\eqref{levy_multiscaling} is satisfied for some $\mu=1+\alpha_{_{\rm exp}}$
and LW modeling approach is the most reasonable one; (ii) other multiscaling laws,
biscaling or not, emerge and neither LW nor HEBP cannot be applied.

%
%

\section*{Acknowledgments}

\noindent
\small
This research is supported by the Basque Government through the BERC 2014--2017
and BERC 2018--2021 programs,  and by the Spanish Ministry of Economy and
Competitiveness MINECO through BCAM Severo Ochoa excellence accreditation
SEV--2013--0323 and through project MTM2016--76016--R ''MIP''. 
VS acknowledges BCAM, Bilbao, for the financial support to her internship
research
period during which she developed her Master Thesis research useful for her
Master degree in Physics at University of Bologna, 
and SV acknowledges the University of Bologna for the financial support through
the ''Marco Polo Programme'' for her PhD research period abroad spent at
BCAM, Bilbao, useful for her PhD degree in Physics at University of Bologna.
PP acknowledges financial  support from Bizkaia Talent and European Commission
through COFUND scheme, 2015 Financial Aid  Program for Researchers, project
number AYD--000--252 hosted at BCAM, Bilbao.
Authors would also like to acknowledge the usage of the
cluster computational facilities of the BCAM--Basque Centre for Applied mathematics
of Bilbao.

\normalsize

\appendix

\section{Numerical algorithms}
\label{num_algo}

In the following we shortly describe the algorithms used for numerical
evaluations and stochastic simulations.

\vspace{.2cm}
\begin{itemize}
\item[(1)]
Langevin equation ~\eqref{eq:langevin_new} for $(x,v)$ is numerically
integrated by using a Euler scheme.
\item[(2)]
  The random drawings from the PDF $f(D)$ are carried out by means of the
Chambers--Mallows--Stuck algorithm for the simulation of extremal L\'evy
densities  $L_{\alpha}^{-\alpha}$ in the interval $0<\alpha<1$
\cite{chambers_etal-jasa-1976,weron-spl-1996,pagnini_etal-fcaa-2016}.
\item[(3)]
To get random drawings from $g(\tau)$, we first obtained a numerical PDF
by applying the Chambers--Mallows--Stuck algorithm to the L\'evy extremal
density $L_{\eta}^{-\eta}$ with very high statistics, and then we computed the
numerical histogram and the function $g(\tau)$ using Eq. ~\eqref{eq:tau_distr}.
This numerical PDF $g(\tau)$ is then used to draw number by the cumulative
function method [see also details in Ref. \cite{vitali_jrsi18}].
\item[(4)]
The HEBP-PDF with maximum diffusivity $D_{\rm max}$, given by Eq. 
~\eqref{dmax_model}, is evaluated by the numerical calculation of the
integral. This is done by applying a composite trapezoidal rule with
sufficiently small step $\delta D$.
\item[(5)]
In the stochastic simulations we used $\alpha=3/2$, $\eta=1/2$
(so that the MSD is $\langle x^2 \rangle \propto t^{3/2}$),
initial conditions $x_{i,0} = 0$ and $v_{i,0} = 0$.
$10^4$ trajectories were simulated, being each trajectory obtained as
an average over a set of $N=9600$ couples $(\tau,D)$ drawn from
$g(\tau)$ and $f(D)$ (for a total of $9.6\cdot10^7$ drawn couples  $(\tau,D)$).
The programs for the simulations were written in the c++ language (Debian
gcc 4.9) and Python 2.7.
\end{itemize}

\vspace{.2cm}
\noindent
A final observation is in order.
A single trajectory of the HEBP can be interpreted as resulting from the
superposed effects of $D$ and, especially, $\tau$ populations.
We applied this approach in performing the numerical simulations.
In particular, for a fixed diffusivity value $D$, the HEBP model becomes
a Gaussian process with long-range time correlation and variance
given by Eqs. ~\eqref{corr_free} and ~\eqref{var_superdiff_1}, respectively.
This is a consequence of the Central Limit Theorem, as the process results
from a linear superposition of independent Gaussian processes.\\
Alternatively, the model can be interpreted as a set of different
particles whose motion $(x(t),v(t))$ is driven by heterogeneous parameters
$(\tau,D)$. Thus, the average motion can be derived as an average over
this ensemble of trajectories.\\
To the goal of stochastic simulations the two approaches are mathematically
equivalent. The two physical interpretations of the trajectories are different,
but the possibility of guessing the correct interpretation depend on the
possibility of observing a sufficiently large number of single trajectories
with high precision, so that both single particle and ensemble statistics
can be computed (similarly to the direct evaluation of ergodic condition).

\section{From HEBPs to RSGPs}
\label{rsgp-ggou}

\noindent
The dynamical equations of  an ensemble of particles moving in a viscous fluid
and having homogeneous parameters $\tau$ and $D$ are:
\begin{eqnarray}
  &&\frac{dx}{dt} = v\ ,
  \nonumber \\
  \ \nonumber \\
  && \frac{dv}{dt} =-\frac{1}{\tau} v(t) + \sqrt{2 D}\, \xi(t)\ .
 \nonumber
\end{eqnarray}
The Langevin equation for $v(t)$ is formally the same as Eq. ~\eqref{eq:langevin_new}, but without the index $i$.
Let us assume that some randomness is assumed for the the diffusivity $D$ and
let us write the process $(x(t),v(t))$ as:
\be
x(t) = \sqrt{2D}\, x_G(t)\ ;\quad v(t) = \sqrt{2D}\, v_G(t)\ .
\nonumber
\ee
by substituting in the previous equations, we get a Langevin equation with unitary noise
intensity:
%
\begin{eqnarray}
  &&\frac{dx_G}{dt} = v_G\ ,
  \nonumber \\
  \ \nonumber \\
  && \frac{dv_G}{dt} =-\frac{1}{\tau} v_G(t) + \xi(t)\ .
 \label{lang_diff1}
\end{eqnarray}
It is well known that the solution for the PDF $p(x_G,v_G,t)$ and, consequently, for the
marginal PDFs of $x_G$ and $v_G$, are Gaussian.
Thus, the process $(x_G(t),v_G(t))$ is a Gaussian process.

\noindent
This proves that the process $(x(t),v(t))$ derived from the randomization of
the velocity diffusivity $D$ is equivalent to a RSGP with random
amplitude given by $\sqrt{2 D}$, while the Gaussian process
is ``generated'' by a Langevin equation with unitary noise intensity.

\noindent
When a population of relaxation times $\tau$ with a complex PDF $g(\tau)$ is considered,
the randomized Langevin equation ~\eqref{lang_diff1} still gives a Gaussian process
but, depending on $g(\tau)$, the global correlation function can be non-exponential
and, thus, diffusion can be non-standard.
In particular, when $g(\tau)$ is given by Eq.  ~\eqref{eq:tau_distr}, the correlation function
is long-range (i.e., slow power-law decaying) and anomalous diffusion emerges.

%
%


\section*{References}
%
%

\end{document}